\documentclass[aps,twocolumn,showpacs,nofootinbib,showkeys,superscriptaddress,preprintnumbers,longbibliography,10pt]{revtex4-2}
\usepackage[utf8]{inputenc}

\usepackage{latexsym}
\usepackage{epsfig}
\usepackage{graphicx,subfigure}
\usepackage[normalem]{ulem}
\usepackage{color}
\usepackage{xcolor}
\usepackage{multirow}
\usepackage{hyperref}
\usepackage{numprint}
\usepackage{eucal}
\usepackage{amsmath}
\usepackage{amssymb}
\usepackage{amsmath,amssymb} 
\usepackage{wasysym}

\DeclareFontFamily{U}{mathb}{\hyphenchar\font45}
\DeclareFontShape{U}{mathb}{m}{n}{
      <5> <6> <7> <8> <9> <10> gen * mathb
      <10.95> mathb10 <12> <14.4> <17.28> <20.74> <24.88> mathb12
      }{}

\definecolor{blue}{rgb}{0,120,250}

\usepackage{xcolor}

\newcommand{\ssout}[1]{}
\begin{document}

\def\ga{\mathrel{\raise.3ex\hbox{$>$\kern-.75em\lower1ex\hbox{$\sim$}}}}
\def\la{\mathrel{\raise.3ex\hbox{$<$\kern-.75em\lower1ex\hbox{$\sim$}}}}

\def\be{\begin{equation}}
\def\ee{\end{equation}}
\def\bea{\begin{eqnarray}}
\def\eea{\end{eqnarray}}

\def\betap{\tilde\beta}
\def\del{\delta_{\rm PBH}^{\rm local}}
\def\Msun{M_\odot}

\newcommand{\dd}{\mathrm{d}} 
\newcommand{\Mpl}{M_P} 
\newcommand{\mpl}{m_\mathrm{pl}} 

\newcommand{\CHECK}[1]{{\color{red}~\textsf{#1}}}

\title{Large curvature fluctuations from no-scale supergravity  \\ with  a spectator field  }

\author{Ioanna D. Stamou}
\affiliation{Service de Physique Th\'eorique, Universit\'e Libre de Bruxelles (ULB), Boulevard du Triomphe, CP225, B-1050 Brussels, Belgium}

\date{\today}
\begin{abstract}

We investigate the large curvature perturbations which can lead to the formation of primordial black holes (PBHs) in the context of no-scale supergravity. Our study does not depend on any exotic scenario, such as scalar potentials with inflection points or bulks, and aims to avoid the fine-tuning of model parameters to achieve the formation of PBHs. This formation relies on the quantum fluctuations of a light spectator stochastic field after the inflationary period. Our analysis is based on the $\mathrm{SU(2,1)/SU(2)\times U(1)}$ symmetry, considering both the inflaton and the spectator field. Specifically, we examine existing no-scale models with Starobinsky-like scalar potentials that are consistent with observable constraints on inflation from measurements of the cosmic microwave background (CMB). These models involve two chiral fields: the inflaton and the modulus field. We propose a novel role for the modulus field as a spectator field, responsible for generating PBHs. Our hypothesis suggests that while the inflaton field satisfies the  CMB constraints of inflation, it is the modulus field acting as the spectator that leads to large curvature perturbations, capable to explain the production of PBHs. Additionally, we prioritize retaining the inflationary constraints from the CMB through the consideration of spectator fluctuations.  
 Therefore, by exploring the relationship between these fields within the framework of the $\mathrm{SU(2,1)/SU(2)\times U(1)}$ symmetry, our  aim is  to unveil their implications for the formation of PBHs.

\end{abstract}

\maketitle

\section{Introduction}

The enigma surrounding the origin of dark matter remains one of the most pressing challenges in modern cosmology. A hypothesis gaining interest is that primordial black holes (PBHs) contribute significantly to the dark matter content. This idea has been supported from various observations, including the groundbreaking detection of gravitational waves emitted by a binary black hole merger  by Ligo/VIRGO/Kagra (LVK) collaboration \cite{Abbott:2016blz,Abbott:2017vtc,Abbott:2017gyy,Abbott:2017oio,Abbott:2016nmj}.
Numerous other studies have also provided  evidence, such as investigations into the size and mass-to-light ratio of ultra-faint dwarf galaxies, identification of several microlensing candidates, analysis of spatial correlations in source-subtracted cosmic infrared and X-ray backgrounds, and the observation of supermassive black holes at high redshifts ~\cite{Aggarwal:2018mgp,Silk_2017,Simon_2019,Cappelluti_2013,Kashlinsky:2016sdv,Kashlinsky:2018mnu,Carr:2020xqk,Bean:2002kx,Clesse:2015wea,Carr:2019kxo,Carr:2018rid,Carr:2023tpt,Huang:2023chx}.
Hence,  these studies  have triggered the theoretical investigations exploring the  existence of PBHs and their  intriguing link to dark matter \cite{Clesse:2017bsw,Garcia-Bellido:2017imq,Carr:2020gox,Escriva:2022bwe,Belotsky:2014kca}.

Inflationary models within the framework of supersymmetry have gained prominence due to their ability to address the energy scale requirements of inflation, which are typically smaller than the Planck scale \cite{Ellis:1982a,Ellis:2013nxa}.
 They offer advantages by circumventing problematic features encountered in other models. For instance, they address the challenge of achieving a vanishing cosmological constant at the classical level without resorting to excessive fine-tuning \cite{Cremer:1983}. Additionally, no-scale supergravity tackles the $\eta$ problem, which arises in attempts to explain the smallness of the inflaton mass compared to the Planck scale \cite{Cremer:1983,Ellislahanas:1984,Elliskounas:1984a,Elliskounas:1984b,Lahanas:1986}. Hence,  the combination of  supergravity, and specifically, the merits of no-scale supergravity, provides a compelling framework for studying inflationary models that effectively address the energy scale requirements and other challenges encountered in cosmology.

Supergravity models have been proposed in the literature in order to explain the production of PBHs \cite{Dalianis:2018frf,Dalianis:2019asr,Gao:2018pvq,Aldabergenov:2022rfc,Aldabergenov:2020bpt,Dalianis:2021iig,Aoki:2022bvj,CrispimRomao:2023fij,Kawai:2022emp}. These  models often involve modified versions of supergravity models that incorporate extra features such as inflection points or bulks. Additionally, no-scale models have been previously studied in order to explain the production and  fractional abundance of PBHs as candidates for dark matter \cite{Nanopoulos:2020nnh,Stamou:2021qdk,Wu:2021zta,Spanos:2022euu}. These models provide significant results in explaining a substantial fraction of dark matter in the Universe while respecting the constraints of inflation reported by the Planck collaboration \cite{Planck:2018vyg}. However, they typically rely on additional features in the effective scalar potential, which require substantial fine-tuning of the underlying parameters.

In this study, we propose an intriguing alternative scenario within the framework of no-scale supergravity, where the desired features can be achieved without the requirement of  fine-tuning. Our approach is guided by the symmetries proposed in Refs. \cite{Ellis:20181,Ellis:2013nxa}, which are based on a non-compact coset SU(2,1)/SU(2)$\times$U(1).
In particular, no-scale models have previously been   studied and  they can be  in accordance with constraints of inflation from  cosmic microwave background (CMB) anisotropies  measurements \cite{Ellis:2013xoa}. In order to reach these observables, there is the  need of introducing one additional chiral field. Hence,  one of these fields  acts as the inflaton  and the other one  is considered as the modulus field. We propose that the modulus field can play the role of a spectator field and despite the fact that it does not contribute to the inflation, it is able to play an important role after inflation and lead to the formation of PBHs.
 Therefore, within this framework of no scale, we consider the presence of two chiral fields, with one field acting as the inflaton and the other as a spectator field.

The central concept behind the generation of   PBHs through a spectator field lies in the quantum stochastic fluctuations experienced by the field. These fluctuations cause the field to acquire distinct mean values within various  Hubble patches. Given the immense number of these patches, it is inevitable that some of them will exhibit the necessary conditions for the spectator field to possess the precise value required for subsequent quantum fluctuations to engender significant curvature fluctuations across horizon-sized regions during PBHs formation.
In essence, the existence of these specific patches triggers an additional expansion of the Universe, resulting in the amplification of fluctuations and ultimately leading to the formation of PBHs. The analysis of this mechanism has been previously  discussed in Ref\cite{Carr:2019hud}.  Similar models with a curvaton field have been proposed  \cite{Kohri:2012yw}. 

In our analysis we can derive  an extra expansion of the Universe using the $\delta \rm N$ formalism for the aforementioned no-scale  supergravity model. 
The outcomes of our study reveal the existence of  non-Gaussian tails in the distribution of induced curvature perturbations. These non-Gaussian tails lead to the production of PBHs.  Furthermore, these results can have implications for the GW background reported by NANOGrav. 
 Finally, our findings align with the constraints on inflation obtained from measurements of CMB.

The structure of this paper is organized as follows.  In Section \ref{No-scale framework} we provide an overview of the fundamental principles of the no-scale theory that are needed in our analysis. In Section \ref{No-scale with a spectator field} we show how a spectator in the framework of no-scale supergravity can lead to large fluctuations at small scales.  In Section \ref{PBHs production} we present the production of PBHs for the aforementioned fluctuations. In Section \ref{Observational constraints of inflation and fine-tuning Discussion} we discuss the fine-tuning of this model.  In Section \ref{alternative} we briefly propose an  alternative  supergravity model motivated from $\alpha$-attractors models and other  no-scale supergravity models. Lastly, in Section \ref{Conclusion}, we  present our concluding remarks.

\section{No-scale framework}
\label{No-scale framework}
In this section, we describe the key characteristics of no-scale supergravity essential for our analysis.
  Firstly, we consider the $\rm{N=1}$ supergravity action  as follows:
\begin{equation}
\label{01}
\mathcal{S}= \int d^4 x \sqrt{-g} \left( K_{i \bar{j}} \partial_{\mu} \Phi^i \partial ^{\mu} \bar{\Phi} ^{\bar{j}}-V(\Phi)  \right),
\end{equation}
 where   $\mathrm{\Phi^i}$  denotes the chiral fields and $\mathrm{ \bar{\Phi} ^{\bar{j}}}$ the conjugate fields.  $\mathrm{K}$  is the K\"ahler potential, which is a real function and $\mathrm{K_{i \bar{j}}}$ is given by:
\begin{equation*}
K_{i \bar{j}}(\Phi,\bar{\Phi})= \frac{\partial^2 K}{\partial \Phi^ i \partial \bar{\Phi}^{\bar {j}} }\, .
\end{equation*}
We work in Planck units ($\mathrm{M_P=1}$) with $\mathrm{M_P}$ is the reduced mass Planck. 
 

The scalar potential $\mathrm{V}$ can be expressed in terms of the Kähler potential $\mathrm{K}$ and the superpotential $\mathrm{W}$ as:
\begin{equation}
V= e^{K}\Big( K^{i \bar{j}} \mathcal{D} i W \mathcal{D}{\bar {j}} \bar{W}- 3|W|^2\Big)
+\frac{\tilde g^2}{2}(K^i {\mathcal{T}}^a \Phi_{i})^2.
\label{eq:potential}
\end{equation}
In this equation, $\mathrm{K^{i \bar{j}}}$ is the inverse of the Kähler metric  and  $\mathcal{D}$ represents the covariant derivative, which are defined as:
\begin{equation}
\begin{split}
& \mathcal{D}_i W  \equiv  \partial_i W + K_i W , \\
& \mathcal{D}^i W   \equiv  \partial^i W - K^i W  \, .
\end{split}
\label{eq:derivofkahler}
\end{equation}
The term of Eq.(\ref{eq:potential}) proportional to $\mathrm{e^{K}}$ is the $\mathrm{F}$-term of the potential. The last term of  this equation proportional to $\mathrm{\tilde g^2}$ corresponds to the $\mathrm{D}$-term potential, which is zero in this case since the chiral fields are gauge singlets.
By  this expression, the scalar potential $\mathrm{V}$ is derived from Eq.(\ref{eq:potential}) for a given K\"ahler potential and superpotential. 

A basic example of no-scale  model is represented by the following equation:
\begin{equation}
\label{eq:kahler_1field}
K=-3 \ln( T+ \bar{T} )\, ,
\end{equation} 
\noindent
where  $\rm{T}$ is a single chiral field. From  Eq.~(\ref{eq:potential}) the term $-3|W|^2$ vanishes due to the identity \cite{Cremer:1983}: 
\begin{equation}
\label{nscont}
K^{T\bar{T}}K_TK_{\bar{T}}=3 \, . 
\end{equation}
Therefore,  in the framework of no-scale theory, the scalar potential is identically zero.
The K\"ahler potential in the no-scale theory given in Eq. (\ref{eq:kahler_1field}) can be identified with the coset SU(1,1)/U(1), which is invariant under the SU(1,1) group of isometric transformation \cite{Elliskounas:1984a}:
\begin{equation}
T \rightarrow \frac{\alpha T+i \beta}{i \gamma T +\delta}, \quad \alpha,\beta,\gamma,\delta \in \rm \mathbb{R}\quad \text{with}  \quad \alpha\delta+\beta\gamma =1 \, .
\end{equation}
This identification allows for a number of attractive features in the theory, including the prediction of a naturally vanishing cosmological constant, a mechanism for generating a large hierarchy between the electroweak and Planck scales, and a natural explanation for the smallness of neutrino masses\cite{Cremer:1983,Cremmer:1983bf,Ellis:1985yc,Ellis:1984bs,Ellis:2020lnc,Ellis:2019opr,Ellis:2016ipm}.

The SU(1,1)/U(1) symmetry becomes more apparent when we introduce the complex field  $y$ as \cite{Elliskounas:1984b,Ellis:2013nxa}:
\begin{equation}
T =\frac{1}{2} \left (   \frac{1-y/ \sqrt{3}}{1+ y \sqrt{3}}  \right).
\end{equation}
Then the K\"ahler potential takes the form:
\begin{equation}
K=-3\ln\left (   1-\frac{y^2}{3}  \right ) .
\end{equation}
The SU(1,1)/U(1) coset space can be parameterized by the $2 \times 2$ complex matrix ${U^\dag }JU$ with:
\begin{equation}
U=\begin{bmatrix}
A & B \\
\bar{B} & \bar{A} 
\end{bmatrix} 
\quad \text{where} \quad
J=\begin{bmatrix}
1 & 0 \\
0 & -1 
\end{bmatrix} \text{and}\quad  A,B  \in \rm \mathbb{C} .
\end{equation}
In this new basis the Lagrangian is invariant under the transformation law:
\begin{equation*}
y \rightarrow  \frac{A y +B}{\bar{B}y +\bar{A}} \quad \text{with} \quad |A|^2-|B|^2=1.
\end{equation*}

The minimal no-scale SU(1,1)/U(1) model lacks the ability to generate the Starobinsky effective scalar potential due to the absence of superpotentials that support it \cite{Ellis:2013nxa}. To overcome this limitation, an extension of the SU(1,1)/U(1) coset to the SU(2,1)/SU(2)$\times$U(1) coset is considered.
 In this new coset the  K\"ahler potential is given from the equation \cite{Elliskounas:1984b}:
\begin{equation}
K=-3 \ln \left( T +\bar{T} -\frac{\phi \bar{\phi}}{3}\right) \, ,
\label{eq:kahler_potential}
\end{equation}
where $\mathrm{T}$ and $\mathrm{\phi}$  are two chiral superfields.  The equivalent form of the K\"ahler potential in the  more revealing basis of two chiral fields $\mathrm{(y_1,y_2)}$, as discussed in Ref. \cite{Ellis:20181} and it is given as follows:
\begin{equation}
K=-3\ln\left(1-\frac{|y_1|^2}{3}-\frac{|y_2|^2}{3}\right).
\label{eq:kahler}
\end{equation}
These complex fields $(y_1,y_2)$ are related to $(T,\varphi)$  by the following expressions:
 \begin{equation}
 y_1=\Big(\frac{2\varphi}{1+2T}\Big) ,\quad y_2=\sqrt{3}\Big(\frac{1-2T}{1+2T}\Big)
 \label{k2(2)}
 \end{equation}
 \noindent
 and the inverse relations by:
  \begin{equation}
 T=\frac{1}{2}\Big(\frac{1 -y_2 / \sqrt{3}}{1 +y_2 / \sqrt{3}}\Big) ,\quad \phi=\Big(\frac{y_1}{1 +y_2 / \sqrt{3}}\Big) \, .
 \label{k2(3)}
 \end{equation}
The superpotential transforms as:
 \begin{equation}
 W(T,\varphi) \rightarrow \bar{W}(y_1,y_2)= (1+y_2 /\sqrt{3})^3W.
  \label{k2(4)}
 \end{equation}
The K\"ahler potential presented in Eq. (\ref{eq:kahler}) is embedded within the SU(2,1)/SU(2)$\times$U(1) coset manifold. This coset can be parametrized by the following matrix \cite{Ellis:20181}\cite{Ellis:2013nxa}:
\begin{gather}
U
 =
  \begin{bmatrix}
   \alpha &
  \beta & 0 \\
   -\beta^* & \alpha^* & 0 \\
   0& 0& 1 
   \end{bmatrix}
\end{gather}
\noindent
where $\alpha,\beta \in\mathbb{C} $ and $|\alpha|^2+|\beta|^2=1$.  By this matrix and the analysis shown in Ref.~\cite{Ellis:20181} we can obtain the following transformation laws for the fields:
\begin{equation}
y_1 \rightarrow \alpha y_1 + \beta y_2, \quad y_2 \rightarrow - \beta^* y_1 +\alpha ^* y_2 \, .
\label{eq:tranformationy12}
\end{equation}
By applying the transformation described in Eq. (\ref{eq:tranformationy12}) to the K\"ahler potential given in Eq. (\ref{eq:kahler}), we observe that it remains invariant. However, the superpotential undergoes non-trivial modifications. Nevertheless, by carefully choosing the parameters $\alpha$ and $\beta$, it is indeed possible to obtain an invariant form of the superpotential. This intriguing feature allows us to explore the symmetries and transformations within the theory, revealing deeper insights into its underlying structure.

A carefully chosen form of the superpotential can yield an effective scalar potential reminiscent of the Starobinsky model. One notable example is the Cecotti superpotential, presented as follows:
 \begin{equation}
W= \sqrt{3}m \phi \left(T- \frac{1}{2}\right) \, ,
\label{eq:w_cecotti}
\end{equation} 
where the field $\rm{T}$ plays the role of inflaton and $\rm{\phi}$ is the modulus field.  The Eq.(\ref{eq:w_cecotti}) can be written in the equivalent form in the basis $\rm{y_1,y_2}$ as \cite{Ellis:2013nxa}:
\begin{equation}
W=m\Big( -y_1y_2 +\frac{y_2y_1^2}{\sqrt{3}}\Big ),
\label{k2(6)}
\end{equation}
where $\rm{ y_1}$ is the inflaton and $\rm{y_2}$ is the modulus field.  
The   Starobinsky-like potential can be derived  from Eq.(\ref{eq:potential}) in the direction $\rm{y_2=Imy_1=0}$.  One should also take into account to fix the non canonical kinetic term in the Lagrangian. 
Hence we have the field redefinition as:
\begin{equation}
\varphi \rightarrow \sqrt{3} \tanh \left( \frac{\varphi}{\sqrt{6}}\right) \, ,
\label{eq:canonical kinetic term1}
\end{equation}
where $\rm{\varphi=Re(y_1)}$.

An alternative superpotential that leads to the Starobinsky scalar potential is found in the simplest globally symmetric model known as the Wess-Zumino model \cite{Ellis:2013xoa}. This model involves a single chiral superfield, denoted as $\phi$, and is characterized by a mass term, denoted as $\hat{\mu}$, and a trilinear coupling, denoted as $\lambda$. The superpotential takes the form:
\begin{equation}
W=\frac{\hat{\mu}}{2}\phi^2-\frac{\lambda}{3}\phi^3.
\label{eq:wess-zumino}
\end{equation}
By considering the field $\mathrm{\phi}$ as the inflaton field and the field $\mathrm{T}$ as the modulus field, and setting $\mathrm{\lambda/\mu = 1/3}$ and $\mathrm{\hat{\mu}=\mu\sqrt{1/3}}$, one can identify a specific direction in field space where $\mathrm{T= \bar{T}=c}$ and $\rm{Im(\phi)=0}$. In this direction, the scalar potential can be reduced to the Starobinsky potential. 
The equivalent form of this superpotential in the basis $\rm{(y_1,y_2)}$ is given as \cite{Ellis:2013nxa}:
\begin{equation}
W=\frac{\hat\mu}{2} \Big(y_1^2 +\frac{y_1^2 y_2}{\sqrt{3}}\Big) -\lambda \frac{y_1^3}{3} \, .
\label{k2(5)}
\end{equation} 
Here $\rm{y_1}$  represents the inflaton and $\rm{y_2}$ is the modulus field, and in order to derive the Starobinsky potential we need to consider the field redefinition from Eq(\ref{eq:canonical kinetic term1}). 

 Finally,  additional superpotentials capable of yielding Starobinsky-like potentials include the following:
\begin{equation}
W=m y_2(1-e^{-y_1})(3-y_1^2)
\end{equation}
and
\begin{equation}
W=m y_2 y_1(3-y_1^2), 
\end{equation}
where   this class of models encompasses  the asymmetric T-model 
$\alpha$-attractor and the assymetric  E-model respectively \cite{Kallosh:2013xya,Linde:2018hmx}. We assume the inflationary direction $\rm{y_2=Im( y_1)=0}$ and $\rm{\varphi =Re(y_1)}$. As previously mentioned, it is necessary to address the non-canonical kinetic term in the Lagrangian, as outlined in Eq.~(\ref{eq:canonical kinetic term1}). We need to remark here that this direction is not always stabilized. However an elegant way to stabilize it can be given by adding an extra term in the K\"ahler potential \cite{Kallosh:2013yoa,Kallosh:2014rga,Kallosh:2010xz}, which is given as:
\begin{equation}
K=-3\ln \Big( 1-\frac{|y_1|^2+|y_2|^2}{3} +\frac{g|y_2|^4}{3-|y_1|^2}   \Big).
\label{6}
\end{equation}
The extra g term does not affect the resulting scalar potential in the inflationary direction.

Therefore,  in the  literature on no-scale models that give rise to Starobinsky-like scalar potentials, two chiral fields have been commonly used: the inflaton and the modulus field. In this study, we retain the properties of the inflaton field that are favoured by the constraints of inflation. Additionally, we propose that the modulus field can play the role of spectator field, responsible for the generation of large fluctuations. Therefore, we hypothesize that while the inflaton field satisfies the CMB constraints of inflation, it is the modulus field, which is regarded as  the spectator,  that facilitates the formation of PBHs.

\section{No-scale with a spectator field}
\label{No-scale with a spectator field}

The key idea behind generating PBHs with a spectator field is the quantum stochastic fluctuations that lead to different mean values in various patches. With a large number of patches, some will meet the conditions for the curvature fluctuations during PBHs formation\cite{Stamou:2023vft,Carr:2019hud,Stamou:2023vwz}.
 In our study, we incorporate this mechanism into a no-scale supergravity model.

In the framwork of no-scale $\mathrm{ SU(2,1)/SU(2)\times U(1)}$ theory we consider the K\"ahler potential given in Eq.(\ref{eq:kahler}).
\begin{equation}
K=-3\ln\left(1-\frac{|\Phi|^2}{3}-\frac{|S|^2}{3}\right), 
\label{eq:kahler2}
\end{equation}
where we assume that the field $\mathrm{\Phi}$ is the inflaton and the $\mathrm{S}$ is the spectator.  However, as we will notice later it is possible to exchange the role of these two  field and obtain the same results. 
The superpotential is given by 
\begin{equation}
W=W_\text{inf}+W_\text{spect}
\label{eq:w_all26}
\end{equation}
or more specifically,
\begin{equation}
\begin{split}
W=&M_\text{inf} S(3-\Phi^2)\left( 1-e^{b_\text{inf}\Phi} \right)^2+\\& M_\text{S} \Phi (3-S^2)(1-e^{b_\text{S}S})^2 \, .
\end{split}
\label{eq:general_superpotential}
\end{equation}
 Here, $M_\text{inf}$ and $M_\text{S}$ represent mass scales associated with the inflationary and spectator sectors, respectively, while $b_\text{inf}$ and $b_\text{S}$ are  parameters modulating the field decay in each sector.  
  We assume that these two terms operate at different scales of the Universe's expansion, as indicated by:
\begin{equation}
M_\text{inf} \gg M_\text{S} \, .
\label{eq:limit}
\end{equation}
 The stabilization of this system is studied in Appendix \ref{appendix}.%

Our proposed model respects the transformation laws of the  coset SU(2,1)/SU(2)$\times$U(1) described in Eq.(\ref{eq:tranformationy12}). Specifically if we introduce the
\begin{center}
 $\Phi \rightarrow -\Phi $ and  $S \rightarrow -S$, 
\end{center}
we derive the identical scalar potential from Eq.\ref{eq:potential} .
To achieve this, it is imperative to properly address the non-canonical kinetic terms in the Lagrangian. For example,  if we assume the direction $\rm{Im\Phi=S=0}$ and $\rm{\varphi=Re\Phi}$ , we introduce a redefinition for the canonical kinetic term as follows:
\begin{equation}
\varphi \rightarrow -\sqrt{3} \tanh \left( \frac{\varphi}{\sqrt{6}}\right).
\label{eq:canonical kinetic minos}
\end{equation}

Moreover, if we implement the fields' transformation in  our case described in Eqs. (\ref{eq:kahler2}) and (\ref{eq:general_superpotential}) as
\begin{center}
 $\Phi \rightarrow S $ and  $S \rightarrow \Phi$, 
\end{center}
we can notice that effectively it changes  the roles of the inflaton and spectator fields.   Concurrently, we adjust the values of $\rm{b_i}$ and $\rm{M_i}$, where  i=\{inf,S\}, to reflect the changed dynamics and to meet the requirements for the inflationary and post-inflationary evolution of the Universe. 
As the Universe cools down from the hot, dense state,  it undergoes phase transitions. Such transitions could naturally lead to changes in $\rm{b_i}$ and $\rm{M_i}$, reflecting the symmetry properties of different phases\cite{Ellis:2017jcp,Ellis:2020lnc,Cheng:2018ajh}. 

To sum up, our analysis underscores the significance of the coset SU(2,1)/SU(2)×U(1) symmetry in shaping the foundational aspects of our model.   In the following, we will explore the dynamics of this no-scale model, focusing on its behavior during inflation, where the inflaton plays a dominant role, and post-inflation, where the spectator field takes precedence.

\subsection{No-scale model during inflation}

One of the intriguing consequences of the no-scale theory is its ability to naturally generate the Starobinsky potential.   We assume the general K\"ahler potential given in Eq.(\ref{eq:kahler2}) and the superpotential  given in Eq. (\ref{eq:general_superpotential}) where, as we mentioned, that  the $\rm{\Phi}$ plays the role of inflaton. Hence we have the following inflationary direction
\begin{equation}
S=Im\Phi=0, \quad Re\Phi=\varphi \, .
\label{eq:inflationary_direction}
\end{equation}
From Eq. (\ref{eq:potential}) we find that 
 the effective scalar potential for the inflaton is given as
\begin{equation}
V_\text{inf}= M_1^2 \left( 1 -\exp \left[ \sqrt{3}b_\text{inf}  \tanh\left( \frac{\varphi}{\sqrt{6}}\right) \right] \right)^4 \, ,
\label{eq:staro}
\end{equation}
where $M_1=3M_\text{inf}$. 
In order to evaluate  the scalar potential given in (\ref{eq:staro}), it is essential to apply the appropriate field redefinition for the canonical kinetic term in the Lagrangian, as specified in Eq. (\ref{eq:canonical kinetic term1}). 
\begin{figure}[ht!]
\centering
\includegraphics[width=80mm,height=55mm]{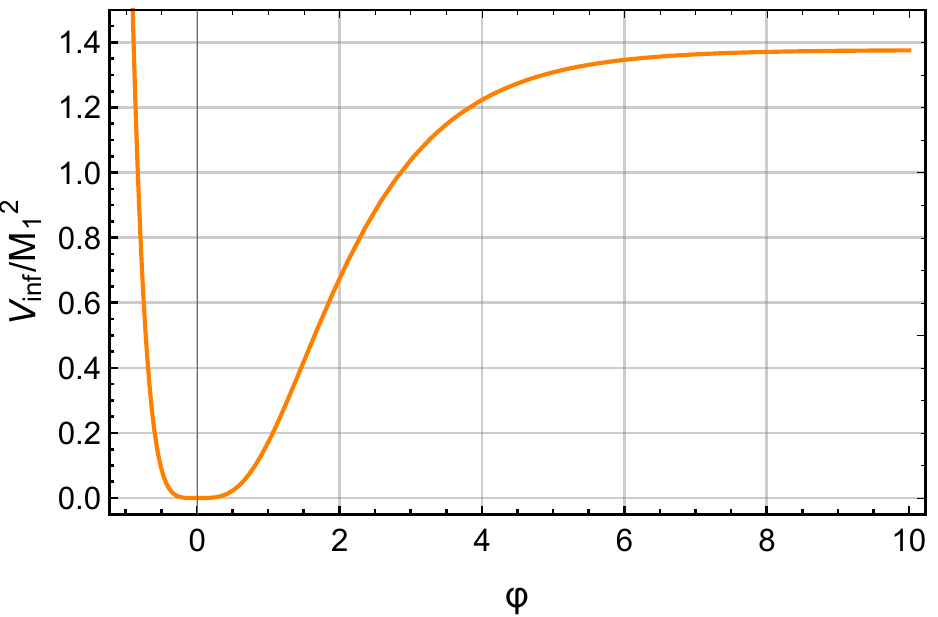}
\caption{ The effective scalar potential  for the inflaton field,  Eq.(\ref{eq:staro}), assuming $\rm{b_\text{inf}=- M_P}$. }
\label{f0}
\end{figure}
In Fig~\ref{f0} we depict the potential for Eq.~(\ref{eq:staro}). %

The evolution of the inflaton field $\rm{\varphi}$ in terms of cosmic time is given as
\begin{equation}
\begin{split}
\ddot \varphi&+3 H \dot \varphi +\frac{\partial V_{inf}}{\partial \varphi} =0,\\
\end{split}
\label{eq:inflaton_evol}
\end{equation}
where  dots denote derivatives in cosmic time. $\mathrm{H}$ is the Hubble parameter which is defined as follows
\begin{equation}
\begin{split}
H =\dot N&= \sqrt{\frac{\rho}{3 }}  
\end{split} 
\label{eq:field.t1}
\end{equation}
and $\rho$ denotes the energy density.

  The slow roll parameters $\mathrm{\epsilon_1}$  and $\mathrm{\epsilon_2}$ defined as 
\begin{equation}
\epsilon_1 \equiv -\frac{d \ln H}{dN},\quad \epsilon_2 \equiv \frac{d \ln \epsilon_1}{dN}.
\end{equation}
The prediction of $\mathrm{n_s}$ and $\mathrm{r}$ can  be approximated from:
\begin{equation}
n_{\rm s} \simeq	 1-2\epsilon_{1*} -\epsilon_{2*}, \quad r=16\epsilon_{1*}\, .
\label{eq.ns}
\end{equation}
Here, the asterisk (*) denotes the values evaluated at the pivot scale $k_* = 0.05 {\rm Mpc}^{-1}$ .

 If we assume as initial condition $\mathrm{\varphi_{ic}=6}$ $\rm{M_P}$ and we solve the Eq.(\ref{eq:inflaton_evol}) we can obtain the following prediction of spectral index $\mathrm{n_s}$ and tensor-to-scalar ratio $\mathrm{r}$:
\begin{equation}
n_s= 0.96907, \quad r=0.002716,
\end{equation}
and   the corresponding slow-roll parameters are given as:
\begin{equation}
 \epsilon_{1*}= 0.00017, \quad \epsilon_{2*}=0.03058\, .
 \label{eq:inflationary_parameter}   
\end{equation}

The power spectrum at the pivot scale  is given from:
\begin{equation}
A_{\rm s}=2.1 \times 10^{-9}\simeq \frac{H^2_*}{8\pi^2 \epsilon_1 M_{\rm P}^2}\, .
\label{eq:amplitude}
\end{equation}
If we consider Eqs.~(\ref{eq:inflationary_parameter}) and  (\ref{eq:amplitude}), the value of the Hubble parameter is given by:
\begin{equation}
H_{*}=5.36 \times 10^{-6} M_P\, .
\label{eq:hcmb_value}
\end{equation}

Therefore, in the inflationary direction Eq.(\ref{eq:inflationary_direction}) we have a model which is in complete consistence with the observable constraints of inflation by Planck collaboration \cite{Planck:2018vyg}.  The remarkable agreement between the predictions of the no-scale Starobinsky model and observational data not only  support to the theory but also highlights the importance  of  study no-scale supergravity models.

\subsection{No-scale model after inflation}

Following the end of inflation, the dynamics of the spectator field come into play, leading to its eventual dominance in the energy density of the Universe. This field  triggers an additional phase of expansion, during which the dynamics of the spectator field play a crucial role in shaping the subsequent evolution of the Universe.

\begin{figure}[ht!]
\centering
\includegraphics[width=85mm,height=55mm]{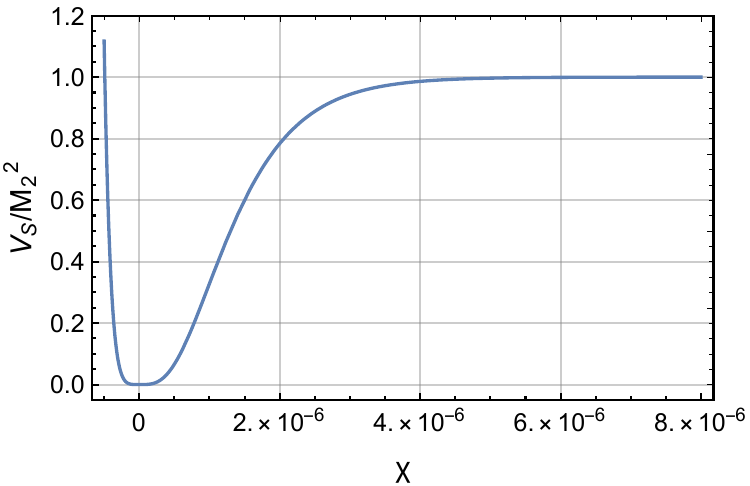}
\caption{ The effective scalar potential for the spectator field, Eq.~(\ref{eq:pot.spec}), 
assuming $\rm{ \mathrm{b_S= - 2 \times 10^{6}} M_P}$. }
\label{eq:field}
\label{f2}
\end{figure}

We consider that the field $\mathrm{S}$ in Eqs.~(\ref{eq:kahler2}) and (\ref{eq:general_superpotential}) plays the role of the spectator.
On scales larger than the Hubble radius, the spectator field remains constant. However, at smaller scales, quantum fluctuations cause the field to undergo fluctuations around its mean value. These mean values can vary across different patches that extend beyond the Hubble radius. If the mean value of the field exhibits sufficient flatness, it will assume a position of dominance in the Universe's density and initiate a brief secondary inflationary period. This secondary phase gives rise to the generation of significant fluctuations in the spectator field. In the following, we will delve into the formalism employed to characterize the origin of these fluctuations in detail.

We examine the Kähler potential and superpotential as specified in Eqs.~(\ref{eq:kahler2}) and (\ref{eq:general_superpotential}). For the spectator field, we adopt the following direction:
\begin{equation}
\Phi=ImS=0, \quad ReS=\chi.
\label{eq:spectator_direction}
\end{equation}
Using Eq.~(\ref{eq:potential}) to calculate the scalar potential
we find that
\begin{equation}
V_S= M_2^2\left( 1- \exp\left[- \sqrt{3}b_S  \tanh \left({\chi}/{\sqrt{6}}\right)\right] \right)^4
\label{eq:pot.spec}
\end{equation}
where $M_2=3M_s$ and we have fixed the non-canonical kinetic term as before
\begin{equation}
\chi \rightarrow \sqrt{3} \tanh \left( \frac{\chi}{\sqrt{6}}\right).
\end{equation}
The parameter $\rm{M_S}$ signifies the scale of the post-inflationary era and does not influence the result of the field's evolution. The parameter $\rm{b_S}$ is approximately defined as $\rm{b_S \approx }\rm{({-10}/{H_{*}}})$, with a detailed examination of its value presented in Ref.~\cite{Stamou:2023vft}. In Fig.~\ref{f2}, we illustrate the potential, highlighting the chosen parameter value in the figure's label. It is evident that our analysis operates within the regime of small field values $\rm{\chi \ll 1}$, ensuring the applicability of Eq.~(\ref{eq:pot.spec}).

The dynamical behavior of the spectator field $\rm {\chi}$ in cosmic time is governed by the equation:
\begin{equation}
\ddot{\chi} + 3H\dot{\chi} + \frac{\partial V_S}{\partial \chi} = 0 \, .
\label{eq:eoms_for_spectator}
\end{equation}
For the energy density  $\rm{\rho}$, we consider the following equation:
\begin{equation}
\rho = \rho_{\text{rad}} e^{-4N} + V(\chi)\, ,
\end{equation}
under the premise that the PBHs are formed during a radiation-dominated epoch. We specify $\rm{\rho_{\text{rad}} = 10V(\chi_{\text{ic}})}$, where $\rm{\chi_{\text{ic}}}$ signifies the initial conditions of the field.

\begin{figure}[h!]
\centering
\includegraphics[width=85mm]{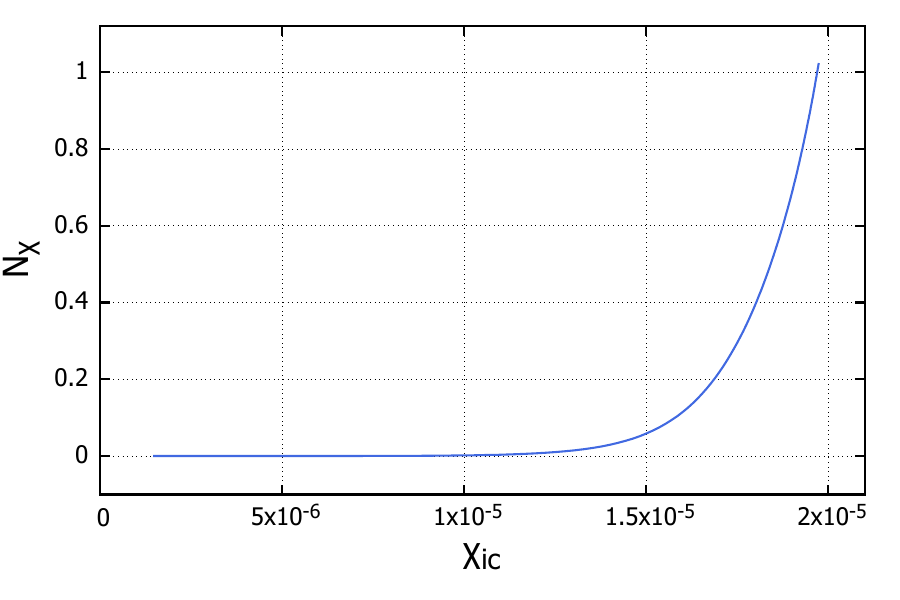}
\caption{ The number of e-fold when the spectator field is dominant. $\rm{\chi_{ic}}$ represents the initial conditions   for the evolution of the Eq.(\ref{eq:eoms_for_spectator}).}
\label{f3}
\end{figure}

\begin{figure}[h!]
\centering
\includegraphics[width=85mm,height=65mm]{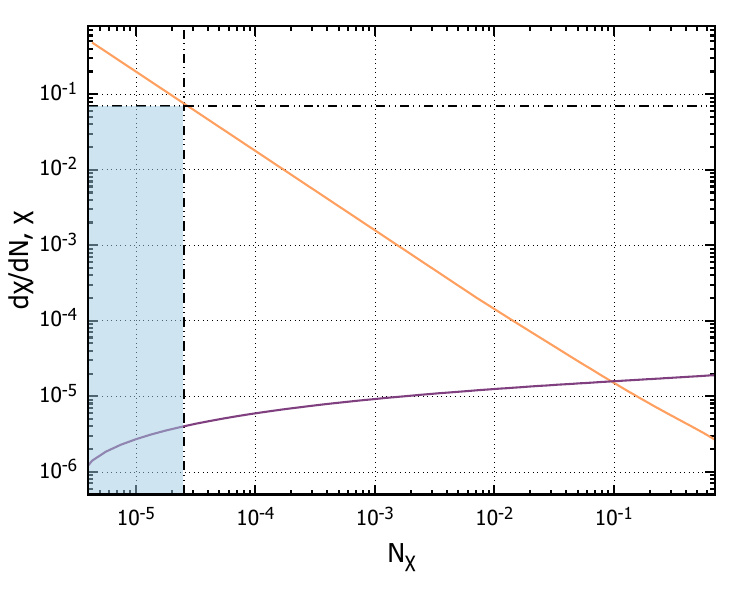}
\caption{ The evolution of the spectator field (purple line) after inflation  and its derivative (orange line). Blue region corresponds to the acceptable range of the  mean value of the field.}
\label{f4}
\end{figure}

The results of the numerical evaluation of Eq.(\ref{eq:field}) using the $\rm{\delta N}$  formalism are depicted in Fig.\ref{f3}, where we observe the additional expansion of the universe that we require. In Fig.~\ref{f4}, the purple line represents the evolution of the field, and the orange line corresponds to its derivative.
It is essential to highlight that the amplitude of the spectator field in this model is consistent with the constraints from the power spectrum at CMB scales. Specifically, by computing the power spectrum amplitude with the equation:
\begin{equation}
P_{R(s)}=\frac{H^2_{CMB}}{4 \pi^2}\left( \frac{d \chi}{dN}\right)^{-2} \Big|_{\chi=\langle \chi\rangle},
\label{eq:as_spe}
\end{equation}
where $\langle \chi\rangle$ is the mean value of the field, we demand to obtain values $\mathrm{P_{R(s)}	\ll
2.1\times 10^{-9}}$. 
 In Fig.~\ref{f3}, the region where the derivative of the field complies with the power spectrum at CMB scales is highlighted in blue. Requiring 
$\rm{P_R<2.1 \times 10^{-10}}$ as an upper limit   and   referring to Eq.(\ref{eq:hcmb_value}), we deduce that the mean value of the field must be less than $\rm{4 \times \mathrm{10^{-6}}M_P}$. Dash-dotted lines on the graph indicate the maximum permissible value to avoid violating the CMB power spectrum constraints. Thus, this model not only potentially contributes to a significant abundance of PBHs, capable of accounting for the dark matter in the Universe, as discussed later but also adheres to the observational constraints at CMB scales.


\section{Curvature perturbations and PBHs production}
\label{PBHs production}

The behavior of the spectator field involves natural randomness, which can be described using a Fokker-Planck equation \cite{Hardwick:2017fjo}. This equation describes the evolution of the probability density distribution associated with the field's fluctuations over time. By solving the Fokker-Planck equation, we can obtain the statistical behavior of the field, allowing us to analyze its properties and study the formation of structures, such as PBHs. 

In our study, we employ the $\rm{\delta N}$ formalism to describe the curvature perturbation $\rm{\zeta}$, which is expressed as:
\begin{equation}
\zeta (x) =  N (\delta \chi_{\rm in} + \delta \chi_{\rm out} + \langle \chi \rangle ) - \langle N\rangle \, , 
\end{equation}
where $\rm{\langle N \rangle}$ represents the average number of e-folds.
 The probability distribution is evaluated from the following integral \cite{Stamou:2023vft,Stamou:2023vwz}
\begin{equation}
P(\zeta_{\rm in}-\zeta_{\rm out}) = \int {\rm d} \delta \chi_{\rm out} P(\delta \chi_{\rm in})  P(\delta \chi_{\rm out}) \left. \frac{{\rm d} \chi}{{\rm d} N}\right|_{\chi_{\rm out}} ~,
\label{eq:prob_dis}
\end{equation}
where $P(\delta \chi_{\rm out})$ and $P(\delta \chi_{\rm in})$ are Gaussian distributions of the fluctuations of the spectator field and  their  corresponding variances come from the inflationary epoch. 
The subscripts "in" and "out" correspond to fluctuations of the field within each patch and around the patches, respectively. 
The evolution of the variance fluctuations $\langle \delta {\chi{\text{out}}}(N_{\text{inf}}) \rangle$ and $\langle \delta {\chi_{\text{in}}}(N_{\text{inf}}) \rangle$ during inflation can be approximated by the following relations.\cite{Stamou:2023vft}:

\begin{equation}
\langle \delta \chi_{\text{out}}^2(N_{\text{inf}}) \rangle \simeq \frac{{H_{{*}}^2}}{{8 \pi^2 \epsilon_{1*}}}\left( 1 - e^{-2\epsilon_{1*} N_{\text{inf}}}\right)
\end{equation}

\begin{equation}
\langle \delta \chi_{\text{in}}^2(N_{\text{inf}}) \rangle \simeq \frac{{H_{*}^2}}{{4 \pi^2 }} \exp\left(-2\frac{\epsilon_{1*}}{\epsilon_{2*}} [e^{\epsilon_{2*} N_{\text{inf}}}-1]\right).
\end{equation}
The values of the slow-roll parameters are given in Eqs. (\ref{eq:inflationary_parameter}) and  (\ref{eq:hcmb_value}).  For more details about the methodology see Refs.\cite{Stamou:2023vft}.


 \begin{figure}[h!]
\centering
\includegraphics[width=95mm]{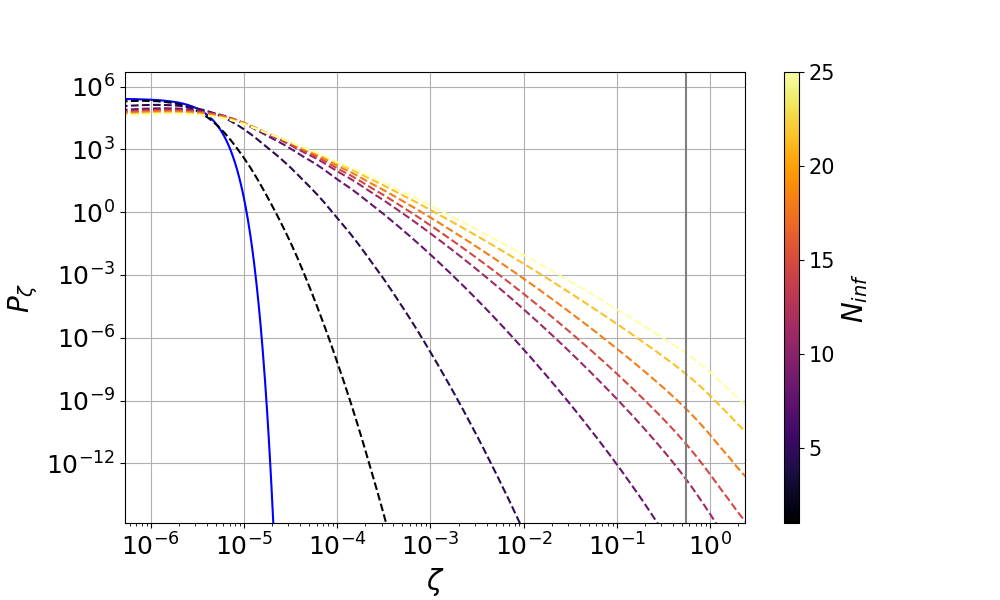}
\caption{ The probability distribution by Eq.(\ref{eq:prob_dis}) assuming $\langle \chi \rangle = 2\times 10^{-6}M_P$.  }
\label{f6}
\end{figure}

 \begin{figure}[h!]
\centering
\includegraphics[width=90mm]{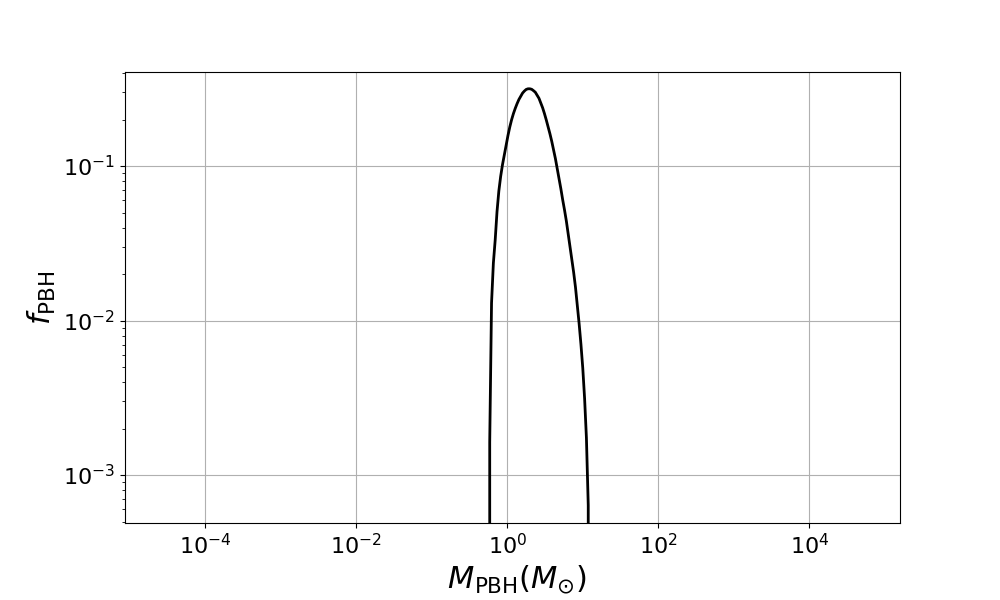}
\caption{ The fractional abundance of PBHs with $\langle \chi \rangle =2 \times 10^{-6}M_P$.    }
\label{f7}
\end{figure}

The formation of PBHs  occurs when a cosmological perturbation collapses to a black hole if its amplitude exceeds a certain threshold value, denoted as $\rm{\zeta_{cr}}$. The threshold value depends on the specific expansion mechanism and properties of the collapsing object, such as its mass, size and  density profile~\cite{Harada:2015yda,Shibata:1999zs,Escriva:2020tak,Musco:2004ak,Musco:2018rwt,Stamou:2023vxu}.  
In Fig.~\ref{f6}, the probability distribution is depicted as a function of  $\rm{\zeta}$, with the gray vertical lines representing the critical threshold $\rm{\zeta_{cr}}$. The blue line represents the corresponding Gaussian distribution with the same inflationary slow -roll parameters of the model given previously.  As the e-folding time increases, we note that the non-Gaussian tails of the distribution widen,  surpassing the $\rm{\zeta_{cr}}$ threshold and hence they  lead to the formation of PBHs.

The probability, which is related to the mass fraction of PBHs,  is evaluated by the integration of the probability distribution described in (\ref{eq:prob_dis}). 
 This probability,    which  depends of the $\mathrm{N_{inf}}$, 
 is connected to the mass fraction of the Universe collapse to PBHs as
\begin{equation}
\beta(M_{\rm PBH}) = \int_{\zeta_{\rm cr}}^{\infty}  P(\zeta) {\rm d} \zeta \,.
\label{eq:beta_derivative}
\end{equation}
The fraction abundance of PBH is given from
\begin{equation}
f_{\rm PBH}(M_{\rm PBH}) \approx 2.4 \beta(M_{\rm PBH}) \left( \frac{2.8 \times 10^{17} M_\odot}{M_{\rm PBH}} \right)^{1/2}~,
\end{equation}
where $\mathrm{ M_{PBH}}$ is the mass fraction of PBHs. More details about the evaluation are shown in Ref.\cite{Carr:2019hud,Stamou:2023vft,Stamou:2023vwz}. 
The fraction abundance of PBHs resulting from our proposed model is illustrated in Fig. \ref{f7}. It is notable that our model can account for a significant fraction of dark matter, offering a compelling explanation. 
Finally, our model can leave an imprint in the GW background, specifically in relation to the observations reported by the NANOGrav collaboration\cite{Kohri:2020qqd,DeLuca:2020agl}.



\section{Fine-tuning Discussion}
\label{Observational constraints of inflation and fine-tuning Discussion}

Inflationary models that facilitate the formation of PBHs often face criticism for the extensive fine-tuning required to amplify the scalar power spectrum adequately, as discussed in \cite{Stamou:2021qdk,Cole:2023wyx}. This section explores the fine-tuning aspects of our model, offering a comparative analysis with other models referenced in the literature.

To address  the fine-tuning concern linked with a given  parameter  of a model $\rm{p}$, we evaluate the quantity $\Delta_{p}$, as introduced in Refs.~\cite{Barbieri:1987fn, Leggett:2014mza}. This quantity is calculated as the maximum value of the logarithmic derivative:
\begin{equation}
\Delta_{p} = \text{Max} \left| \frac{\partial \ln(f)}{\partial \ln(x)} \right| = \text{Max} \left| \frac{\partial f}{\partial x} \cdot \frac{x}{f} \right|,
\label{deltaeenz} 
\end{equation}
where $f(x)$ denotes a general function. High $\Delta_{p}$ values mean the model needs a lot of fine-tuning, showing it's very sensitive to changes in its parameters.

 One can numerically compute Eq.~(\ref{deltaeenz}) by assuming the function $\rm{f}$ represents the fractional abundances of PBHs, $\rm{f_{PBH} (p)}$. In models featuring an inflection point, recent work has indicated that this quantity ranges from $\rm{10^4}$ to $\rm{10^9}$  \cite{Cole:2023wyx}. The specific range varies depending on the model. However, the significant level of fine-tuning required across these scenarios raises concerns regarding the naturalness of such models. Hybrid models have been proposed as a more natural mechanism for generating PBHs. The degree of fine-tuning in these models has been analyzed in \cite{Afzal:2024xci,Spanos:2021hpk,Braglia:2022phb} demonstrating a significant reduction in the parameter described by Eq.~(\ref{deltaeenz}), with values approximately  to $\rm{10^2}$.


In our analysis, the only additional parameter is the  $b_S$ of the potential. This parameter   dependents on the inflationary model we have and especially on the value of the Hubble parameter, $\mathrm{H_{*}}$. Its value is crucial in order to obtain  correct values for the  power spectrum at CMB scales and it does not affect  the fractional abundance.  Other values of $b_S$, which are smaller  than the one provided here (over than one order of magnitude), lead to obtaining significant fraction of PBHs but they affect the in Eq.(\ref{eq:as_spe}).  A detailed discussion about the fine-tuning is provided in Ref.\cite{Stamou:2023vft}. By evaluating the quantity Eq.~ (\ref{deltaeenz}) we find that this takes values of $\rm{\mathcal{O}(1)}$. 
Therefore, this model can indeed avoid the need of fine tuning of the parameters in order to obtain a significant fraction of PBHs. 

As we mentioned before, PBHs from no-scale have previously been   studied in Refs.\cite{Nanopoulos:2020nnh,Spanos:2022euu}.
  The amount of fine-tuning has been computed to be at the range of $\mathrm{10^6}$ \cite{Hertzberg:2017dkh,Stamou:2021qdk}.   Moreover, these models have been criticised from  the evaluation of one-loop corrections  to the large-scale power spectrum
   according to the study of Refs. \cite{Kristiano:2022maq,Fumagalli:2023loc,Choudhury:2023vuj,Choudhury:2023hvf,Bhattacharya:2023ysp,Choudhury:2023rks,Choudhury:2023kdb,Choudhury:2023jlt,Choudhury:2024ybk,Firouzjahi:2023aum,Inomata:2022yte}.
   An other issue, which have these models to deal with, is  the prediction of extremely light PBHs, lighter than the range of the future detection and, hence, their results cannot be tested in the near future. Finally, in these models the condition for obtaining a vanishing cosmological constant given in Eq.(\ref{nscont}) is not conserved as well as the properties of the  SU(2,1)/SU(2)$\times$U(1) coset manifold.

In this proposed model, on the other hand,  it is important to remark that slightly changes of the parameter $b_S$  does not affect the results. Hence this model does not suffer for the  unnatural fine-tuning of the inflection-point models.  
 In addition to that,  in our analysis we use the $\mathrm{\delta N}$ formalism, which evades the problem with the  one-loop correction~\cite{Firouzjahi:2023ahg}. Moreover, the prediction of masses are in the range of  $\mathrm{[10^{0}-10^{2}]}$ $\mathrm{M_\odot}$ which are supported by subsolar mass black hole secondary component, which is considered as GW signal \cite{Morras:2023jvb}. Finally, the properties of the group  SU(2,1)/SU(2)$\times$U(1)  are conserved,   as it is analysed previously.

\section{Alternative scenario in context of supergravity models}
\label{alternative}

\subsection{ An $\alpha$-attractors supergravity model}
Several models can apply  the proposed mechanism of spectator field analyzed in Ref. \cite{Stamou:2023vft} in order to elucidate PBHs production. This section presents an alternative   model for generating substantial curvature perturbations necessary for the formation of PBHs. This alternative scenario is based on superconformal  $\alpha$-attractors supergravity models.

In the framework  of superconformal field theory, the landscape is enriched by symmetries that extend beyond local supersymmetry \cite{Kallosh:2014ona}.  These include Special Conformal Symmetry,
Special Supersymmetry, 
and Weyl Symmetry. 

The alternative model is based on the following choice of K\"ahler potential and superpotential, extensively explored in the literature \cite{Kallosh:2013yoa,Kallosh:2013hoa,Kallosh:2014rga, Kallosh:2010xz,Kallosh:2013xya,Kallosh:2014ona,Kallosh:2010ug}
\begin{equation}
K_2=-3\alpha \ln \Big( 1-\frac{|S|^2+|\Phi|^2}{3} +\frac{g|\Phi|^4}{3-|S|^2}   \Big)
\label{eq:kahler_spec_alpha}
\end{equation}
and 
\begin{equation}
W_2=\Phi M \frac{S}{\sqrt{3}} \left(3- S^2\right)^{(3\alpha -1)/2}.
\label{eq:w_spec_alpa}
\end{equation}
The $\alpha$ term in $\alpha$-attractor models represents a parameter that controls the curvature of the potential in the inflationary landscape, influencing the inflationary trajectory and the resulting predictions for cosmological observables like the spectral index and the tensor-to-scalar ratio.  In our scenario, this term is applied to the spectator field and does not directly impact the inflationary observable constraints.

The potential in the direction of the spectator field  $\rm{\Phi=ImS=0}$ and $\rm{ReS=\chi}$ is simply given as
\begin{equation}
V_2(\chi)={M}^2 \tanh^2 \left( \frac{\chi}{\sqrt{6 \alpha}} \right),
\label{eq:pt_spectator}
\end{equation}
where we have fixed the non-canonical kinetic term. Regarding the inflaton field, we consider a model that yields slow-roll parameters and observable constraints predictions akin to those in the previously examined no-scale model. 
. 

\begin{figure}[ht!]
\centering
\includegraphics[width=75mm]{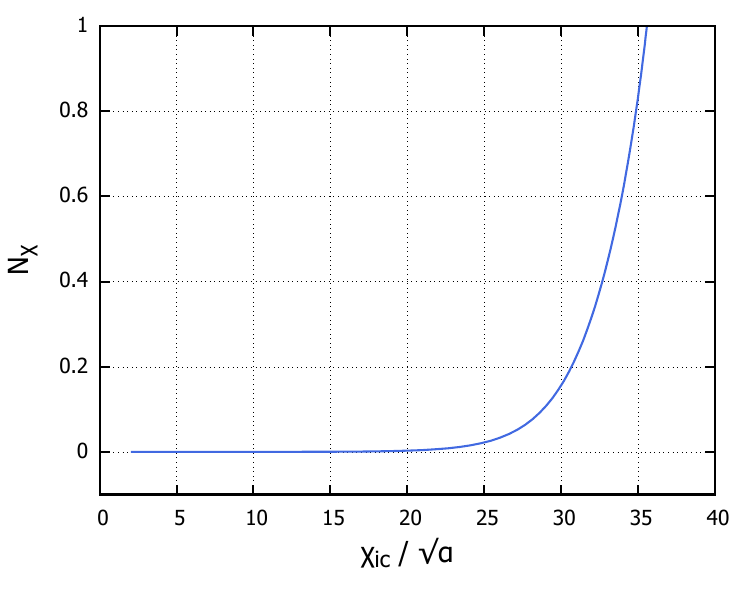}
\caption{ The extra expansion of the Universe assuming $\alpha=10^{-13}$ during the phase when spectator field is dominant.  }
\label{f8}
\end{figure}

\begin{figure}[ht!]
\centering
\includegraphics[width=95mm]{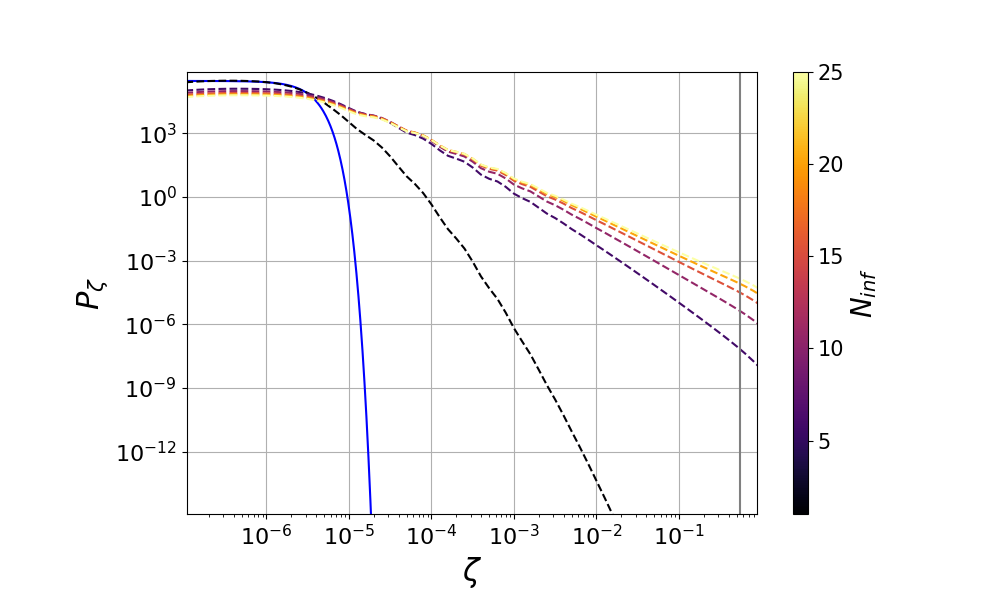}
\caption{ The probability distribution for the case of $\alpha-$attractors.  }
\label{f9}
\end{figure}

We use  the $\rm{\delta}$N formalism to analyze the dynamics of the spectator field, as in the previous case of no-scale model. The resulting additional expansion of the Universe is illustrated in Fig.~\ref{f8}. Thus, this model facilitates the extra  expansion necessary for the formation of PBHs.

In conclusion, we determine the probability distribution using Eq.~(\ref{eq:prob_dis}) and we depict our results  in Fig.~\ref{f9}. The non-Gaussian tails widen as the number of e-folds increases. During certain inflationary e-folds, these tails may surpass the critical threshold, inducing substantial curvature fluctuations sufficient to produce PBHs.

\subsection{ Alternative examples based on no-scale}
In this section, we present alternative superpotential choices that can lead to additional expansion of the Universe, and consequently, a significant fraction of PBHs within the framework of no-scale supergravity.

First, we adopt the Kähler potential from Eq.~(\ref{eq:kahler2}) and consider the following superpotential for the spectator field in Eq.~(\ref{eq:w_all26}):
\begin{equation}
W_S = M_S\left(-S\Phi +\frac{\Phi S^2}{\sqrt{3}}\right)\left(1+c_2e^{-b_2 S^2}S^2\right),
\end{equation}
where S represents the spectator field and $\rm{\Phi}$ is the inflaton and $\rm{c_2}$, $\rm{b_2}$ are parameters.  This choice, explored in \cite{Stamou:2021qdk}, can lead to an inflection point. With an appropriate choice of parameters, this point can be positioned very close to the potential's origin, thereby allowing fluctuations around this point to generate large curvature perturbations, as proposed in Ref.~\cite{Stamou:2023vwz}. Positioning the inflection point near small field values ensures accordance with the power spectrum constraints at CMB scales.

Additionally, we explore the different superpotentials, such as:
\begin{equation}
W = M_{\text{inf}}\left( -\Phi S +\frac{S \Phi^2}{\sqrt{3}} \right) + M_S \Phi S^{b_3}(3- S^2) 
\end{equation}
or 
\begin{equation}
W = M_{\text{inf}}\left( -\Phi S +\frac{S \Phi^2}{\sqrt{3}} \right) +  M_\text{S} \Phi(1-e^{b_\text{S}S})(3-S^2) \, , 
\end{equation}
where $\rm{S}$ is the spectator field, $\rm{\Phi}$ is the inflaton (as before), and $\rm{b_3}$ is a parameter, with the Kähler potential again given by Eq.~(\ref{eq:kahler2}).  We can derive a potential which exhibits a plateau at small field values and it can lead to similar extra expansion of the Universe capable to facilitate the PBHs production. 

Therefore, within the framework of supergravity models, there are multiple avenues to achieve an extra expansion of the Universe, similar to the results of this work.

\section{Conclusion}
\label{Conclusion}
This study delves into the realm of PBHs within the no-scale supergravity framework, offering an alternative avenue avoiding the need of inflection points. 
 Our analysis focuses on the $\mathrm{SU(2,1)/SU(2)\times U(1)}$ symmetry and considers both the inflaton and spectator fields. Specifically, we examine existing no-scale models that generate scalar potentials resembling the Starobinsky model, which typically involve two chiral fields: the inflaton and the modulus field. By interplaying the proposed symmetries from previous works, we introduce these two chiral fields, with the inflaton field driving inflation and the modulus field acting as a spectator throughout the process.

We have developed a mechanism in which a small expansion of the spectator field, approximately one e-fold, plays a crucial role in the formation of PBHs during the radiation epoch. By carefully exploring this concept within the framework of an inflationary model, we have ensured that our predictions adhere to the rigorous observational constraints set by the amplitude of CMB fluctuations on large scales. We evaluate the fractional abundance of PBHs and we find that this model can predict a significant fraction of dark matter in the Universe with large values of mass of PBHs. 

As a next step in our research, we plan to delve into the exploration of the mass distribution of  PBHs within the framework of this model. By analyzing the mass spectrum, we aim to understand better the role that PBHs play as potential dark matter candidates.  Additionally, we intend to examine the compatibility of our model's predictions with existing observational constraints. 

Overall, our study contributes to the growing body of research on PBHs and their connection to dark matter and the evolution of the universe. By offering an
  alternative mechanism within the no-scale supergravity framework and providing insights into the interplay between chiral fields, we deepen our understanding of the intricate dynamics that give rise to PBHs. This research opens up new avenues for exploring the nature of dark matter and the mechanisms that shape the Universe's structure.

\appendix
\section{Study the stabilization }
\label{appendix}
In this appendix, we examine the stabilization of the fields presented in the Kähler potential and superpotential in Eqs.~(\ref{eq:kahler2}) and (\ref{eq:general_superpotential}).

\begin{figure}[h!]
\centering
\includegraphics[width=75mm]{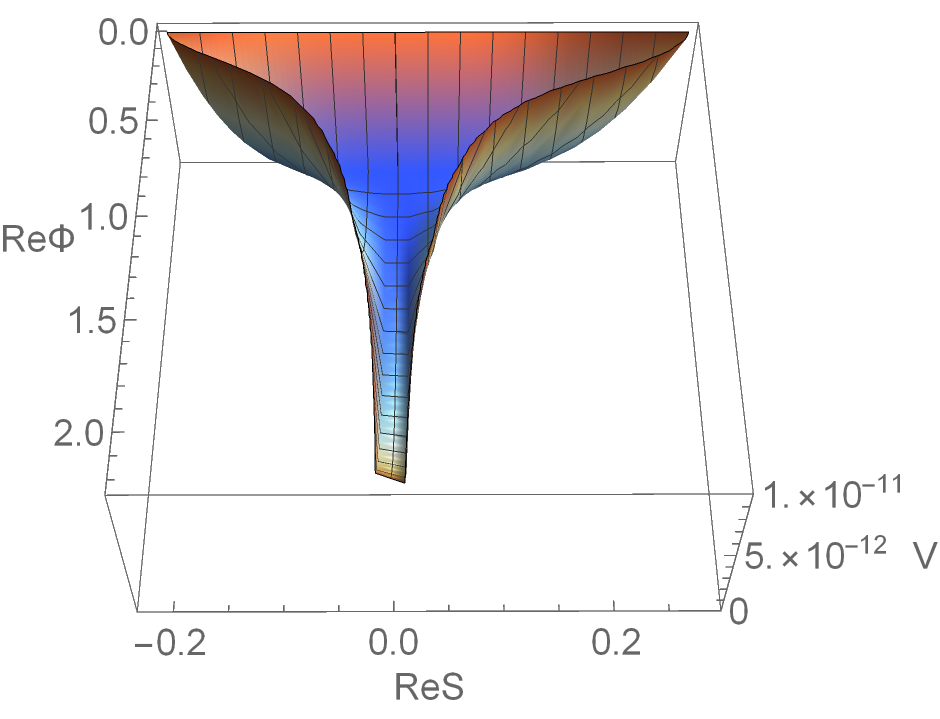}
\caption{ The scalar potential assuming $\rm{\Phi=\bar{\Phi}=Re \Phi}$ and $\rm{\bar S =S= Re S}$ using ~(\ref{eq:kahler2}) and (\ref{eq:general_superpotential}). }
\label{f10}
\end{figure}

\begin{figure}[h!]
\centering
\includegraphics[width=42mm]{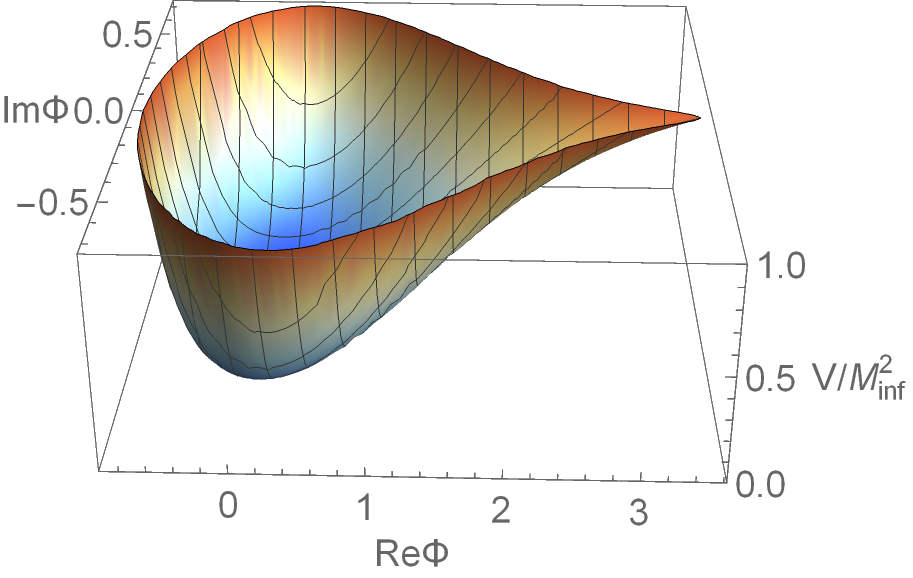}
\includegraphics[width=42mm]{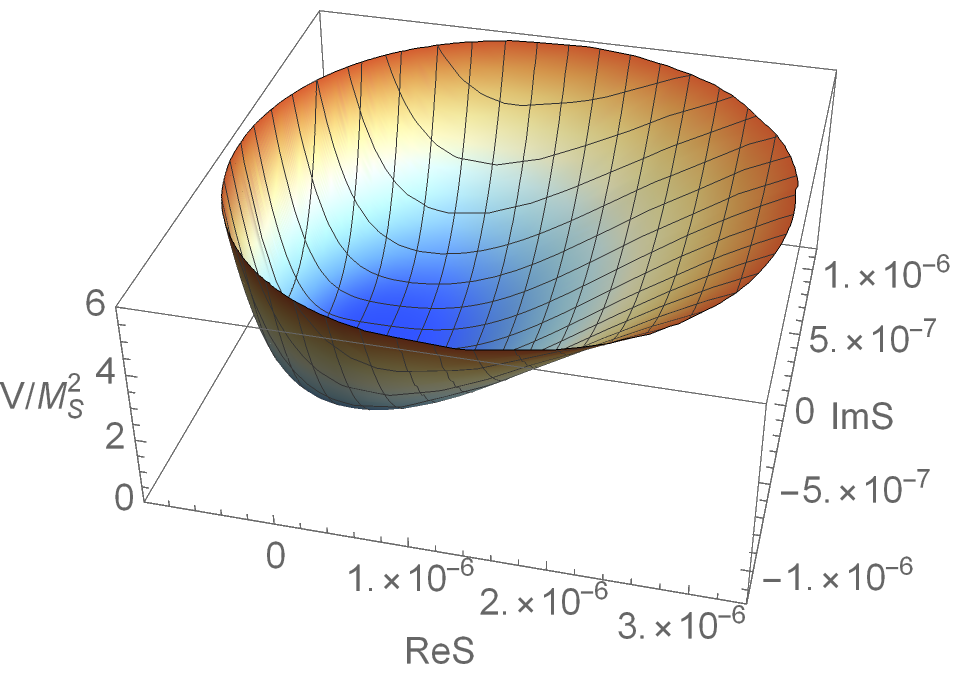}
\caption{ Slices through the effective potential for the model from Eqs.(\ref{eq:kahler2}) and (\ref{eq:general_superpotential}) along the direction $(\rm {Re\Phi, Im\Phi})$ (left plot) and $(\rm {ReS, ImS})$ (right plot). }
\label{f11}
\end{figure}

First, we calculate the squared mass matrix using the method outlined in Ref.~\cite{Ellis:2019bmm} for our proposed model.


\begin{equation}
m^2_{s}=\begin{pmatrix} 
\alpha_{\Phi\bar{\Phi}} & \alpha_{\Phi \bar{S}} \\
\alpha_{ S \bar{\Phi}}   & \alpha_{S \bar{S}}
\end{pmatrix}.
\label{eq:matrix_s}
\end{equation} 
We calculate the  elements of the matrix and we set $\Phi =\bar{\Phi}$ and $S =\bar{S}$ (see Eq.(18) in \cite{Ellis:2019bmm}). 
The elements are:
\begin{equation}
\alpha_{\Phi \bar \Phi} =(K^{-1})^i_k D^k \partial_j V
\end{equation}

\begin{equation}
\alpha_{\Phi \bar S} =(K^{-1})^k_i D_k \partial_j V
\end{equation}

\begin{equation}
\alpha_{ S \bar \Phi} =(K^{-1})^i_k D^k \partial^j V 
\end{equation}

\begin{equation}
\alpha_{ S \bar S} =(K^{-1})^k_i D_k \partial^j V 
\end{equation}
where $\rm{i,j,k}$ denote the fields and the covariant derivatives are given in Eq.(\ref{eq:derivofkahler}). 
 The scalar potential is described in Eq.(\ref{eq:potential}) and it is depicted in Fig.\ref{f10}. 	The inflaton slowly rolls through the valley at the direction of $\rm{S= Im\Phi=0}$ until the end of inflation. Then at values very close to the origin of the potential the spectator field $\rm{S}$ starts to dominate the Universe. In Fig.\ref{f11} we depict various slices of the potential  along the real and  imaginary part of each field.  Our results are similar to those in Ref.\cite{Ellis:2014gxa}

After identifying the matrix elements in Eq.(\ref{eq:matrix_s}), we set the expectation values  $\langle S \rangle= 0$ and $\langle \Phi \rangle=0$. In order to find this values, it becomes necessary to  numerically solve the following system  $\partial {V_j} \big|_{j=\langle j \rangle} = 0$ where j represents the fields.
 Given our selected parameters and within the chosen field region, we find that the mass matrix exhibits positive eigenvalues. This calculation is repeated for the imaginary direction, confirming that, given our parametric selection, the eigenvalues of the matrix  have positive solutions.  However, considering a more extensive range of parameters one need to incorporate the introduction of additional stabilizing terms to the Kähler potential.
Such modifications align with the suggestions in Refs.~\cite{Kallosh:2013yoa,Kallosh:2013lkr,Ellis:2014rxa}, indicating that stabilization is achievable. Incorporating these stabilizer terms is particularly model in the context of a two-field system, suggesting that a more comprehensive analysis is imperative for a generalized scenario.

\section*{Acknowledgement}
We would like to express our gratitude to  S. Clesse, P. Fayet,  M.A. Garcia Garcia \& M. Vanvlasselaer for their fruitful discussions. 

\bibliographystyle{apsrev4-2}

\bibliography{bib.bib} 




\end{document}